\documentclass{aa}  

\usepackage[breaklinks=true,colorlinks,citecolor=blue]{hyperref}
\usepackage{graphicx}
\usepackage{enumerate}
\usepackage{epsfig}
\usepackage{color}
\usepackage{txfonts}
\usepackage{natbib}
\usepackage{multirow}
\usepackage{ulem}

\newcommand{\kmps}{\rm km~s\ensuremath{^{-1} }\,}
\newcommand{\kmskpc}{km~s\ensuremath{^{-1}}~kpc\ensuremath{^{-1} }\,}

\newcommand{\Gaia}{{\it Gaia}\,}

\begin{document}

\title{Hic sunt dracones\thanks{{\it Here be dragons}, a phrase famous in medieval cartography  when dragons and sea monsters were used to designate uncharted and possibly dangerous regions.}: Cartography of the Milky Way spiral arms and bar resonances with Gaia Data Release~2}

\author{S.~Khoperskov$^{1}$, O.~Gerhard$^{1}$, P.~Di Matteo$^{2}$, M.~Haywood$^{2}$,  D.~Katz$^{2}$, S.~Khrapov$^{3}$, A. Khoperskov$^{3}$, M.~Arnaboldi$^{4}$}
\authorrunning{S. Khoperskov et al.}

\institute{$^{1}$Max-Planck-Institut f\"{u}r extraterrestrische Physik, Gie{\ss}enbachstrasse 1, 85748 Garching, Germany \\  $^{2}$GEPI, Observatoire de Paris, PSL Universit{\'e}, CNRS,  5 Place Jules Janssen, 92190 Meudon, France \\ $^{3}$Volgograd State University, 100 Universitetskii prospect, 400062, Volgograd, Russia \\ $^{4}$European Southern Observatory, Karl-Schwarzschild-Str. 2, 85748 Garching, Germany}


\abstract{In this paper we introduce a new method for analysing Milky Way phase-space which allows us to reveal the imprint left by the Milky Way bar and spiral arms on the stars with full phase-space data in \Gaia Data Release 2. The unprecedented quality and extended spatial coverage of these data enable us to discover six prominent stellar density structures in the disc to a distance of $5$~kpc from the Sun. Four of these structures correspond to the spiral arms detected previously in the gas and young stars~(Scutum-Centaurus, Sagittarius, Local and Perseus). The remaining two are associated with the main resonances of the Milky Way bar where corotation is placed at around $6.2$~kpc and the outer Lindblad resonance beyond the Solar radius, at around $9$~kpc. For the first time we provide evidence of the imprint left by spiral arms and resonances in the stellar densities not relying on a specific tracer, through enhancing the signatures left by these asymmetries. Our method offers new avenues for studying how the stellar populations in our Galaxy are shaped.}


\maketitle

\section{Introduction}

A long-standing problem in galactic dynamics is how gravitational instabilities of stellar discs shape the large-scale morphology of galaxies, and in particular their conspicuous spiral arms. While the spectacular images of spiral galaxies mainly reflect young stars formed from interstellar gas in spiral arms, all disc stars can cross the arms and contribute to their overdensities; therefore the entire disc should bear the imprint of its spiral structure. In the Milky Way, the position of our Solar System near the plane of our Galaxy, distance uncertainties, and interstellar extinction have complicated attempts to learn directly about the large-scale disc morphology. With massive stars and pulsars~\citep{1993ApJ...411..674T,2006Sci...311...54X,2018A&A...616L..15X}, HII regions ~\citep{1976A&A....49...57G,2003A&A...397..133R} and masers~\citep{2014ApJ...783..130R,2019arXiv191003357R,2017Sci...358..227S} numerous studies have attempted to delineate the spiral arms of our Galaxy~\citep{2014AJ....148....5V}. However, the structure of the disc in old stars, beyond the solar neighborhood, remains largely {\it terra incognita}. 

The second data release~(DR2) of ESA's \Gaia mission has provided the largest available 6D phase-space~(positions and velocities) dataset for $7.2$ million stars brighter than $G_{\rm RVS} = 12$~mag~\citep{2018A&A...616A..11G}, making possible precise studies of the Milky Way structure and kinematics on large scales. \Gaia~data have already revealed signatures of non-equilibrium and ongoing vertical phase mixing in the Milky Way disc \citep{2018Natur.561..360A}, likely induced by a previous pericentric passage of the Sagittarius dwarf galaxy~\citep{2019MNRAS.485.3134L, 2019MNRAS.486.1167B,2019ApJ...879L..15H} or by internally driven bending waves~\citep{2019A&A...622L...6K,2019MNRAS.484.1050D}. A large number of kinematic arches with various morphologies not known prior to \Gaia were found~\citep{2018A&A...619A..72R,2019arXiv190904949K,2019arXiv190801318M} and large-scale wiggles were discovered in the $V_{\phi}-R$ plane~\citep{2018MNRAS.479L.108K,2018Natur.561..360A}. These features can be interpreted as the signature of the impact of an external perturbation~\citep{2016ApJ...823....4D,2019MNRAS.485.3134L} or as evidence of the spiral structure and the bar of the Milky Way~\citep{2018MNRAS.481.3794H,2019MNRAS.488.3324F}. At the same time, in the Solar vicinity, the Galactic spiral structure is believed to generate various patterns in the phase-space~\citep{2012MNRAS.425.2335S,2013MNRAS.436..101W}, but uncertainties in the location and strength of the spiral arms have so far prevented explaining the observed patterns. In this work, we use the high-quality \Gaia~DR2 sample of stars with radial velocities~(hereafter, GRV2), together with a new method to highlight the stellar density structures, to explore the Milky Way spiral arms without relying on any specific stellar tracer. The outline of the Letter is as follows: in Sect.~\ref{sec::data_method}, we describe data and the method we used; in Sect.~\ref{sec::results}, we examine the phase-space structure of the Milky Way and compare its various features with Milky Way-type galaxy simulations; and finally, in Sect.~\ref{sec::summary}, we summarize the conclusions that can be drawn from our study.

\begin{figure*}
\begin{center}
\includegraphics[width=0.8\hsize]{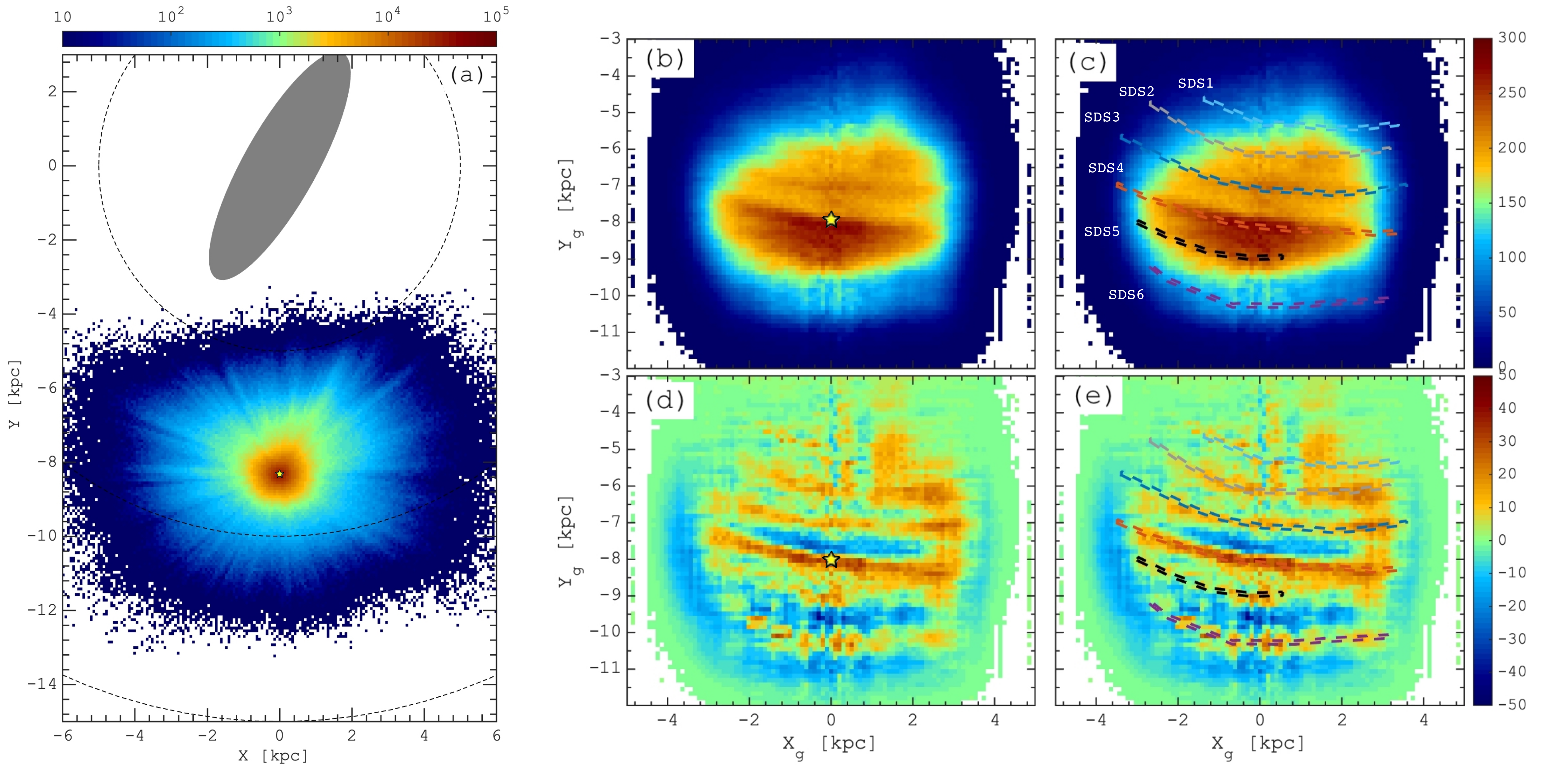}
\caption{ Face-on view on the Milky Way disc with \Gaia DR~2: spatial and guiding coordinates space. {\it (a):} Two-dimensional map in the Galactic plane of the number of \Gaia~DR~2 stars with radial velocities, colour coded by number of stars. The grey oval represents the orientation of the Milky Way bar~\citep{2015MNRAS.450.4050W}, and the Solar position is marked by a yellow star. The stellar density distribution is shaped by the \Gaia~footprint and is heavily influenced by the fading of stars at greater distances and by the high interstellar extinction in the Galactic plane. {\it (b),~(c):}, Two-dimensional map in guiding coordinates space of the mean number of stars over $100$ random samples drawn from the homogenenized spatial density map~$\rm \rho(X_g,Y_g)$. {\it (d),~(e):} Unsharp masking map~(see Eq.~\ref{eq::sharpting}) of the guiding space shows six newly discovered overdensities prominent in a wide area of the Milky Way disc, each containing between 2'000 and 6'000 stars, about 5-12\% of the respective background. This estimate is a lower limit for the amplitude of the overdensities because of the dust extinction in the disc plane  blurring their visible structure. These features are marked in {\it  (c),~(e):}, and are interpreted below as the imprint of the Galactic bar~(corotation and outer Lindblad resonances) and the four-armed spiral structure in the Milky Way. The irregular  vertical features near $X\approx 0$ line are artifacts of the \Gaia~DR~2 footprint in this area.}\label{fig::fig1} 
\end{center}
\end{figure*}

\section{Data and Methods}\label{sec::data_method}
We used \Gaia~DR~2 sources for which the 6D phase space coordinates can be computed, that is all sources with an available 5 parameters astrometric solution~(sky positions, parallaxes, and proper motions) and radial velocities~\citep{2018A&A...616A...1G}. From this sample, we selected stars with positive parallaxes, and relative errors on parallaxes less than $10\%$. Distances were computed by inverting parallaxes.  For calculating positions and velocities in the galactocentric rest-frame, we  assumed an in-plane distance of the Sun from the Galactic centre of $8.19$~kpc~\citep{2018A&A...615L..15G}, a velocity of the Local Standard of Rest, $V_{LSR}=240~\kmps$~\citep{2014ApJ...783..130R}, and a peculiar velocity of the Sun with respect to the LSR, $U_\odot = 11.1$~\kmps, $V_\odot =12.24$~\kmps, $W_\odot=7.25$~\kmps~\citep{2010MNRAS.403.1829S}. After applying these selections, our final sample consists of $\approx 5.3$~M stars, whose density distribution, projected onto the Galactic plane, is shown in Fig.~\ref{fig::fig1}a. As expected, the main shape of the distribution is controlled by the \Gaia selection function, with the density rapidly decreasing with the distance from the Sun. We also investigated the sensitivity of our results on the \Gaia parallax biases~\citep{2019MNRAS.487.3568S} and did not find any systematics changes in the structures we discovered. 

\begin{figure*}[t!]
\begin{center}
\includegraphics[width=0.8\linewidth]{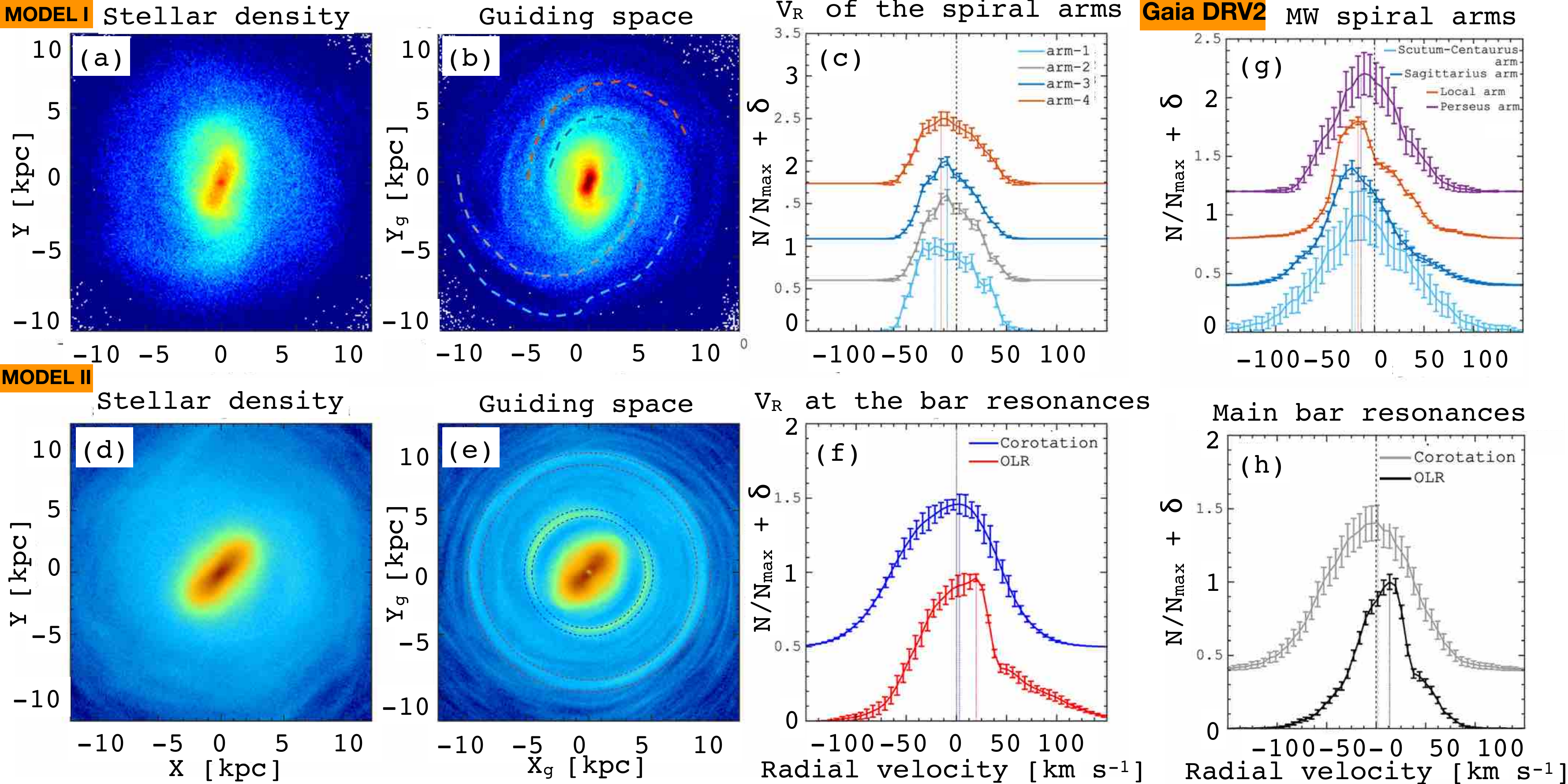}
\caption{Milky Way-type galaxy simulations and kinematic properties of the overdensities in the Milky Way disc. {\it (a)-(c):} Milky Way-type spiral galaxy~(Model I). The stellar surface density in~{\it (a)} shows only marginal evidence for density waves because these are blurred by epicyclic motions. After the transformation to guiding coordinates space~{\it (b)} the model spiral arms are much clearer, and show similar morphology as the gas density (see~Appendix~\ref{app2}). These maps are shown in $\log$-scale. {\it (c)} Stellar radial velocity distributions for the four spiral arms highlighted in panel {\it (b)} where the mean radial velocity within the arms is negative (inwards), exactly as expected from density wave theory~\citep{1969ApJ...155..721L}. {\it (d)-(f):} Milky Way-type barred galaxy~(Model II). Transforming stellar surface density~{\it (e)} to guiding coordinates space~{\it (f)} shows prominent overdensities at the bar's corotation~(CR) and outer Lindblad~(OLR) resonances, as determined from spectral analysis in Appendix~\ref{app2}.  {\it (g)} shows a broad, nearly symmetric radial velocity distribution for stars at CR~(blue), as viewed from the solar position at $\rm (x,y)=(0,-8.2)$ kpc, and and an outward-skewed distribution at OLR~(red). {\it (g), (h):} Kinematics of stars in \Gaia GRV2 associated with the large-scale stellar overdensities~(SDS) detected in guiding coordinates space~(see Fig.~\ref{fig::fig1}). {\it (g):} Radial velocity distribution for four structures (SDS~1,~3,~4,~6) reveals a mean radial motion toward the galactic centre similar, to that seen in the spiral arms of Model~I. {\it (h):} Radial velocity distributions in SDS~2,~5 resemble the distributions at CR and OLR in the in the barred galaxy Model~II. Error bars on all profiles are determined as described in Appendix~\ref{app2}.}\label{fig::fig2}
\end{center}
\end{figure*}

The detection of stellar structures in the Galaxy faces two problems, of different nature. On the one side, the epicyclic motion of stars, that is the radial oscillation around their guiding centres, blurs all kind of morphological features in discs. Such an effect is clearly seen in galactic structures made of multiple stellar populations with different kinematics, making any density structure more evident for younger, dynamically colder stellar populations with lower orbital eccentricities~\citep{2017MNRAS.469.1587D,2018A&A...611L...2K}. On the other side, the \Gaia and \Gaia~RVS limiting magnitudes, the shape of the \Gaia~footprint, and the effect of extinction have the consequence of limiting the accessible portion of the disc to a few kpc from the Sun, with a density of sources that decreases rapidly with distance from the Sun~(see Fig.~\ref{fig::fig1}a). To tackle the first of these two problems,  it is convenient to analyze stellar distribution in what we call the ``guiding coordinates space'', i.e.
\begin{equation}
\rm X_g = - R_g \sin(\phi),\,\, \,
Y_g = - R_g  \cos(\phi)\,,\label{eq::eq1}
\end{equation}
where $\rm R_g = L_z / V_{\rm LSR}$ is the guiding radius normalized to the LSR velocity taken to be $\rm V_{\rm LSR} = 240~\kmps$, $L_z$ is the instantaneous angular momentum of the star $\rm L_z = R \times V_{\phi}$, $R$ its galactocentric distance, $\phi$ is the azimuthal angle around the galactic center clockwise from the direction towards the Sun, and $\rm V_\phi$ its azimuthal velocity in the Galactic plane. By transforming to $\rm (X_g,Y_g)$~or ($\rm R_g, \phi$) phase-space, we change the radial positions of stars without changing their azimuths $\phi$. This allows us to minimize the effect of the radial oscillations of stars around their guiding centres and hence sharpens the face-on features in the stellar density distribution. The guiding space coordinate transformation is based on the approximate conservation of angular momentum along the normal to the disc. This assumption is supported by the fact that the velocity field in the region from outside the bar to beyond the Sun is not far from circular~\citep[][]{2014ApJ...783..130R,2015ApJ...800...83B,2016ARA&A..54..529B}, despite perturbations from the bar and spiral arms, and the vertical disturbances of the disc discovered recently.

In order to reduce the effect of the \Gaia~DR~2 selection function, we apply an approach based on a re-sampling of the face-on density map in the spatial coordinates. First, we introduce a uniform Cartesian grid in spatial coordinates ($\rm X,Y$) with the cell size of $\rm 100\times100$~$\rm pc^2$. Then we populate each grid cell by $100$ randomly selected stars located within this cell. Such a single sampling procedure results in a constant number of stars in the $\rm (X,Y)$-plane on the Cartesian grid. Then by repeating random re-sampling for each cell of the $\rm (X,Y)$-grid we generate $100$ independent selections of stars from the entire \Gaia~DR~2, where each sub-sample contains $100\times100\times100=10^6$ stars. Thanks to the large statistics of the~\Gaia~DR~2 catalogue, we can re-populate our $\rm(X,Y)$-grid by different sub-samples of stars, with repetitions, especially far beyond the Solar vicinity where the number of the observed stars is lower. Then each sub-sample of stars~(with a flat distribution in $\rm (X,Y)$-coordinates) is converted into $\rm (X_g,Y_g)$ guiding phase space map~(see Eq.~\ref{eq::eq1}) where the flat distribution in $\rm(X,Y)$ plane transforms into a map depicting several large-scale features.

\section{Results}\label{sec::results}
Figures~\ref{fig::fig1}~(top) show the mean density map $\rm \rho(X_g,Y_g)$ for $100$ re-sampled selections of stars where several horizontally extended over-density features can be found across the entire disc area covered by \Gaia~DR~2. To sharpen even more the features seen in the $\rm \rho(X_g,Y_g)$ map we produce a residual map by using an unsharp masking approach. In Fig.~\ref{fig::fig1}~(bottom), we show the residual map defined as:
\begin{equation}
\rm \delta \rho(X_g,Y_g) = \rho(X_g,Y_g) - \langle \rho(X_g,Y_g) \rangle\,, \label{eq::sharpting} 
\end{equation}
where $\rm \langle \rho(X_g,Y_g) \rangle$ denotes the mean number density map constructed by convolving the density distribution $\rm \varrho(X_g,Y_g)$ with a 2D Gaussian kernel of width $400$~pc. The combined re-sampling technique and transformation to guiding coordinates reveal several new stellar density structures.

Figure~\ref{fig::fig1}b shows the stellar density now plotted in the guiding coordinates space ($\rm X_g, Y_g$). Large-scale overdensities of length $3-8$~kpc in guiding coordinates cover most of the disc observed by \Gaia. Structures are seen even more clearly once the unsharp masking is applied. The result is presented in Fig.~\ref{fig::fig1}d where six outstanding narrow overdensities of $\approx 5-12$\% amplitude are visible. The most prominent features are seen close to the Solar radius: one crosses the line from the Sun to the Galactic centre at $\rm Y_g\approx-7$~kpc, and another passes through the Solar vicinity at $\rm Y_g\approx-8$~kpc. Two additional structures are detected in the inner galaxy, between $\rm Y_g = -4.5$ and $\rm Y_g=-6$~kpc.  In the outer disc, we find an extended structure between $\rm Y_g = -9$ and $\rm Y_g=-10.5$ kpc and a smaller one just beyond the Solar radius between $\rm Y_g=-8$ to $\rm Y_g=-9$ kpc. Coloured lines in Fig.~\ref{fig::fig1}c,e highlight the main structures that appear in the guiding coordinates space. 

To test possible effects of extinction, we investigated a cross-matched sample between GRV2 and WISE~\citep{2018MNRAS.481.2148W}. We selected stars in this sample with absolute W1 magnitude in [-1,-2], and considered the redder and bluer half in G-W1 colour separately. We found that the redder half of this sample, expected to contain a larger fraction of reddened stars, showed a significantly noisier representation of structures in guiding space than the bluer half, as well as a dominant extinction feature in the direction of the Galactic centre. This suggests that the guiding space structures seen in GRV2 are not the result of sample biases due to extinction, but that, on the contrary, extinction likely reduces the amplitudes of the structures we found. Therefore our estimates of their amplitudes should be considered as lower limits.

\begin{figure*}[t!]
\begin{center}
\includegraphics[width=0.8\hsize]{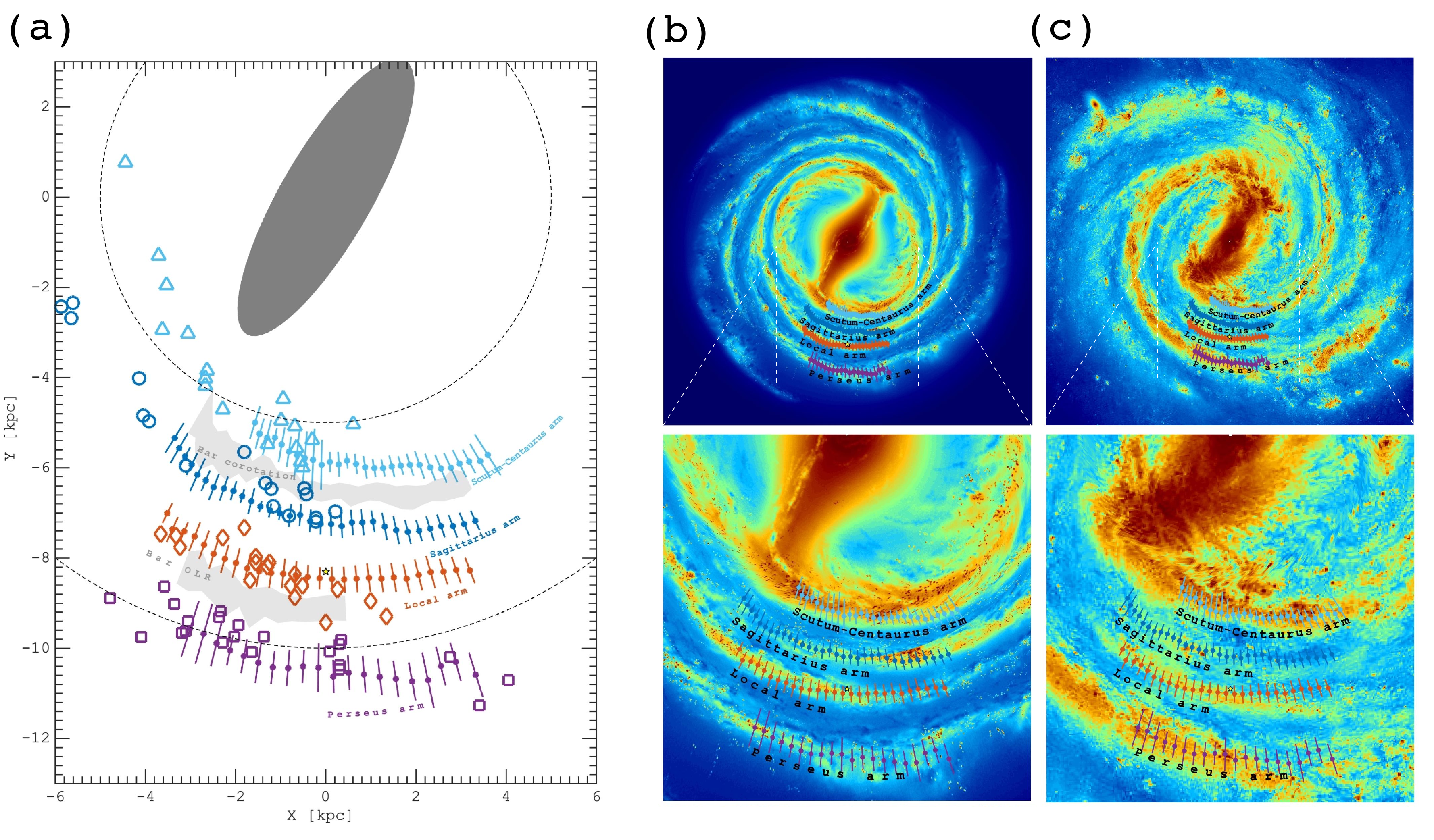} 
\caption{Milky Way spiral arms and bar resonances revealed by \Gaia~DR~2. {\it (a):} \Gaia~DR~2 based spiral arms shown by symbols with error bars while two gray-shaded areas show the location of stars likely associated with co-rotation and OLR of the Milky Way bar -- all structures as identified using guiding coordinate space densities and radial velocity distributions. Large symbols without error bars depict the position of high-mass star-forming regions with trigonometric parallaxes measured by the VLBA~\citep{2014ApJ...783..130R}. These sources trace the gaseous and young star spiral arms in the Milky Way~(Scutum-Centaurus, Sagittarius, Local and Perseus). The grey oval represents the orientation of the Milky Way bar, and the Solar position is marked by a yellow star. {\it (b),~(c):} Two exemplary new artist's views of a Milky Way spiral galaxy including the four-armed tightly-wound spiral arms revealed by \Gaia DR~2. Modified images of original Milky Way galaxy rendering by Robert Hurt and Nick Risinger.}\label{fig::fig3}
\end{center}
\end{figure*}

In order to understand what causes these structures we consider two Milky Way-type galaxy simulations. The first is an $N$-body/SPH simulation of a spiral galaxy~(Model I, see details of the model in Appendix~\ref{app2}) in which the spiral structure in the stellar disc is weak and blurred by the epicyclic motions of the stars (see Fig.~\ref{fig::fig2}a). However, transforming the positions of the simulated stars to guiding coordinates space using Eq.~(\ref{eq::eq1}) allows us to recover narrow spiral arm-like features (see Fig.~\ref{fig::fig2}b) whose morphology turns out in much closer agreement with the gas surface density map than the original stellar density (see Appendix~\ref{app2}). For the simulated stars in each of these features we calculate the distribution of radial velocities (see Fig.~\ref{fig::fig2}c; more details in Appendix~\ref{app2}). In each histogram, we see a systematic inward motion, with the radial velocity distribution skewed towards negative values. This pattern is in agreement with predictions of density wave theory where, within corotation, the spiral arms stars move inwards towards the galactic centre~\citep{1969ApJ...155..721L,2012MNRAS.425.2335S,2014MNRAS.440.2564F,2016MNRAS.461.2383P,2016MNRAS.457.2569M}.

Another non-axisymmetric structure that could generate features in the guiding coordinates space is a stellar bar which gives rise to a number of resonances in the outer disc. In Fig.~\ref{fig::fig2}d-f we show results from a snapshot of an $N$-body simulation of a barred galaxy without prominent spiral structure~(Model II, see details of the model in Appendix~\ref{app2})). Here, two narrow quasi-circular overdensities appear in the guiding coordinates space outside the bar which are absent in the spatial density map. Spectral analysis of the orbits of particles in these overdensities confirms that these features are at the bar's corotation and outer Lindblad (OLR) resonances~(see Appendix~\ref{app2} for details). The observed kinematics of stars at the bar resonances depends on both the bar orientation and the distribution of stellar orbits. In particular, some stars near corotation follow orbits elongated along the bar major axis, showing periodic variations of their radial velocities. The bar OLR is populated by two orbital families with proportions varying along the OLR region. When the bar is oriented similarly as in the Milky Way, and an area of the simulated disc is considered similar to that available with GRV2~($\rm Y<0, |X|<5$~kpc), we found that the OLR region is dominated by $\rm x_1(2)$ orbits~\citep{2019MNRAS.488.3324F} that move outwards. This makes the mean radial velocity as seen from the Sun positive, with a velocity distribution skewed towards positive values. At the same time stars near corotation show different kinematics with more symmetric radial velocity distribution~(see Fig.~\ref{fig::fig2}f and Appendix~\ref{app2} for details). 

We conclude from these theoretical models that both the spiral arms and major bar resonances are imprinted in velocity space. The extended spatial coverage, high sampling, even far from the Solar vicinity, and unprecedented precision of the \Gaia~DR~2 data allow us to see these signatures in the observed velocities. Figure~\ref{fig::fig2}g shows that four of the overdensity structures obtained in Fig.~\ref{fig::fig1} with the guiding space analysis (SDS~1, SDS~3, SDS~4, and SDS~6) have systematically negative radial velocity with a radial velocity distribution skewed towards to negative values, just as the spiral arms in the simulated galaxy of Model~I.  Meanwhile SDS~2 has a wide and more symmetric distribution with only a small mean negative velocity, and stars in SDS~5 tend to move outwards as seen from the Sun, with the velocity distribution skewed towards positive values~(see Fig.~\ref{fig::fig2}h), very similar to the barred galaxy simulation Model~II.

We can also transfer the structures in Fig.~\ref{fig::fig1} back to original Galactic plane coordinates~(see Appendix~\ref{app1}). To place all structures in the Galactic plane, we estimate their mean location by using Gaussian fitting along the galactocentric radial direction with 1D Gaussian models. Figure~\ref{fig::fig3}a shows the inferred locations of the Milky Way spiral arms~(SDS~1, 3, 4, 6). Overplotted are the positions of known high-mass star-forming regions (HMSFRs) with accurate VLBI maser distances~\citep{2014ApJ...783..130R,2016SciA....2E0878X} which trace the Milky Way spiral arms in gas and young stars. These sample the Perseus arm~(purple), the Local arm~(orange), the Sagittarius arm~(blue) and the Scutum-Centaurus arm~(turquoise). The location of these HMSFRs is in excellent agreement with the stellar overdensities we find. The Local arm and the Perseus arm perfectly overlap with SDS~4 and SDS~6, the fourth and the sixth~(from the Galactic centre) stellar density structures  discovered in this study. The star forming regions in the inner disc (associated to the Sagittarius arm and the Scutum-Centaurus arm) are spatially coincident with SDS~1 and SDS~3, the first and third of the discovered stellar structures; a few HMSFRs connecting SDS~1 and SDS~3 also overlap with SDS~2. 

Fitting a logarithmic spiral arm model to the data in Fig.~\ref{fig::fig3}a, we measure the pitch angles~\citep{2012ApJS..199...33D} for the spiral arm structures~(SDS~1, 3, 4, 6); from the inner to outer arm $14.7\pm1.1^\circ$, $10.8\pm0.8^\circ$, $5.8\pm0.7^\circ$, and $7.7\pm0.7^\circ$, respectively.  Therefore, our results suggest that the Milky Way has a four-armed spiral, in agreement with a number of  studies of  MW spiral structure based on different tracers and techniques~\citep{2014AJ....148....5V,2019arXiv191003357R}, although some previous results are in favour of $m=2$ pattern~\citep[e.g.,][]{2012MNRAS.425.2335S}.

Interestingly,  the Local arm, which is the closest spiral arm to the Sun, appears to be a tightly wound structure, located between two primary arms of the Milky Way, Sagittarius and Perseus. This result supports the idea that the Local arm is an intrinsic spiral arm, rather than an inter-arm spur as suggested for a long time~\citep{1980ApJ...242..528E, 2001AJ....122.3017S, 2006ApJ...650..818L,2019ApJ...882...48M}. 

The SDS structures associated with the Milky Way bar resonances~(SDS~2 and SDS~5) are shown as gray-shaded regions in Fig.~\ref{fig::fig3}a where corotation is placed at $R\approx6.2$~kpc and the OLR is right beyond the Solar radius, at $R\approx9.3$~kpc. Note that the derived location of the resonances could correspond to a fraction of the entire resonance structure which is supported by our Model II where both CR and OLR regions cover wide radially-extended areas across the disc~(see Fig.~\ref{fig::run6} in Appendix~\ref{app2} and Fig.~\ref{fig::fig4} in Appendix~\ref{app1}). This issue of completeness of the detected OLR-stars in the disc beyond the Sun will be resolved by applying the approach presented here to stars with radial velocities in the forthcoming \Gaia DR3. Although the Milky Way bar itself is not covered by our GRV2 sample, the location of the bar resonances allows us to estimate its angular rotation speed. For the significantly declining rotation curve of the Milky Way~\citep{2019ApJ...871..120E} we obtain values of $37-42$~\kmskpc in good agreement with recent in situ estimates from Milky Way bar kinematics and dynamics~\citep{2017MNRAS.465.1621P,2019MNRAS.488.4552S,2019arXiv190511404B}.

\section{Conclusions}\label{sec::summary}
Understanding the global structure of our Milky Way, in all its constituents,  is a classical problem of Galactic astronomy.  Mapping the main non-axisymmetric features in the disc, as traced by their stars, is an important advance made possible only by the superb quality of the \Gaia~DR~2 data; so far the Galactic spiral arms had only been detectable with young stellar populations and gas. In this paper we provide a means for identifying stars of spiral arms and bar resonances without relying on any specific stellar tracer, opening up several further avenues towards understanding of the Milky Way.

For the first time, we reveal the imprint left by the Milky Way bar and spiral density waves on the entire disc population, the Gaia Data Release 2 enabling us to recover the prominent four-arm spiral structure of the Galaxy (Scutum-Centaurus, Sagittarius, Local and Perseus spiral arms) to a distance of $5$~kpc from the Sun~(see Fig.~\ref{fig::fig3}b,c). We also independently constrain the Milky Way bar parameters by detecting the prominent signatures of its main resonances which together with properties of spiral arms provide us a much more complete picture of the asymmetries present in the disc of our Galaxy, and never seen before Gaia. The presence of the bar and extended spiral structure evidently induce in-plane phase-space perturbations that imply that the Milky Way stellar disc is out of equilibrium on large scales, and most of the features in the velocity space are related to the presence of these non-axisymmetric structures in the disc. 

In this work we also described a new idea to analyse the Milky Way stellar populations which in combination with \Gaia~data and spectroscopic information~(WEAVE, MOONS, 4MOST) will allow us to study stellar dynamics in the vicinity of the Galactic spiral arms, including radial migration of stars, as well as the influence of the Galactic bar on the disc, and thus the disc's dynamical evolution. It will also shed light on the chemical composition of stars in spiral arms and the role of these structures in shaping the stellar populations over time. 

\begin{acknowledgements}
The authors thank Michael Fall for useful discussions during preparation of the work. This work has made use of data from the European Space Agency (ESA) mission \Gaia (\url{https://www.cosmos.esa.int/gaia}), processed by the \Gaia Data Processing and Analysis Consortium (DPAC, \url{https://www.cosmos.esa.int/web/gaia/dpac/consortium}). Funding for the DPAC has been provided by national institutions, in particular the institutions participating in the \Gaia Multilateral Agreement. PDM and MH thank the ANR (Agence Nationale de la Recherche) for its financial support through the MOD4Gaia project (ANR-15-CE31-0007, P.I.: P. Di Matteo). Development of the code for simulation of stellar-gaseous galactic discs by AK and SK was supported by the Ministry of Science and Higher Education of the Russian Federation (government task No. 2.852.2017/4.6). Numerical simulations were computed with support of the Russian Science Foundation~(project no. 19-72-20089) by using the equipment of the shared research facilities of HPC computing resources at Lomonosov Moscow State University~(project RFMEFI62117X001). This work was also granted access to the HPC resources of CINES under the allocation 2017-040507 (Pi : P. Di Matteo) made by GENCI. 
\end{acknowledgements}

\begin{appendix}

\section{Galaxy models: understanding the Milky Way guiding space features}\label{app2}
\subsection{Model I: spiral galaxy model}\label{app2a}
In order to explore the nature of the large-scale overdensities we discovered in the guiding space of the Milky Way, we present an analysis of two different models of the Milky Way-type galaxies. In particular, to disentangle the perturbations of the guiding phase-space caused by the spiral arms and by the bar, we analyze a model of a multi-arm spiral galaxy~(Model I) and a model with a bar and lack of spiral arms~(Model II).

\begin{figure*}[t!]
\includegraphics[width=1\linewidth]{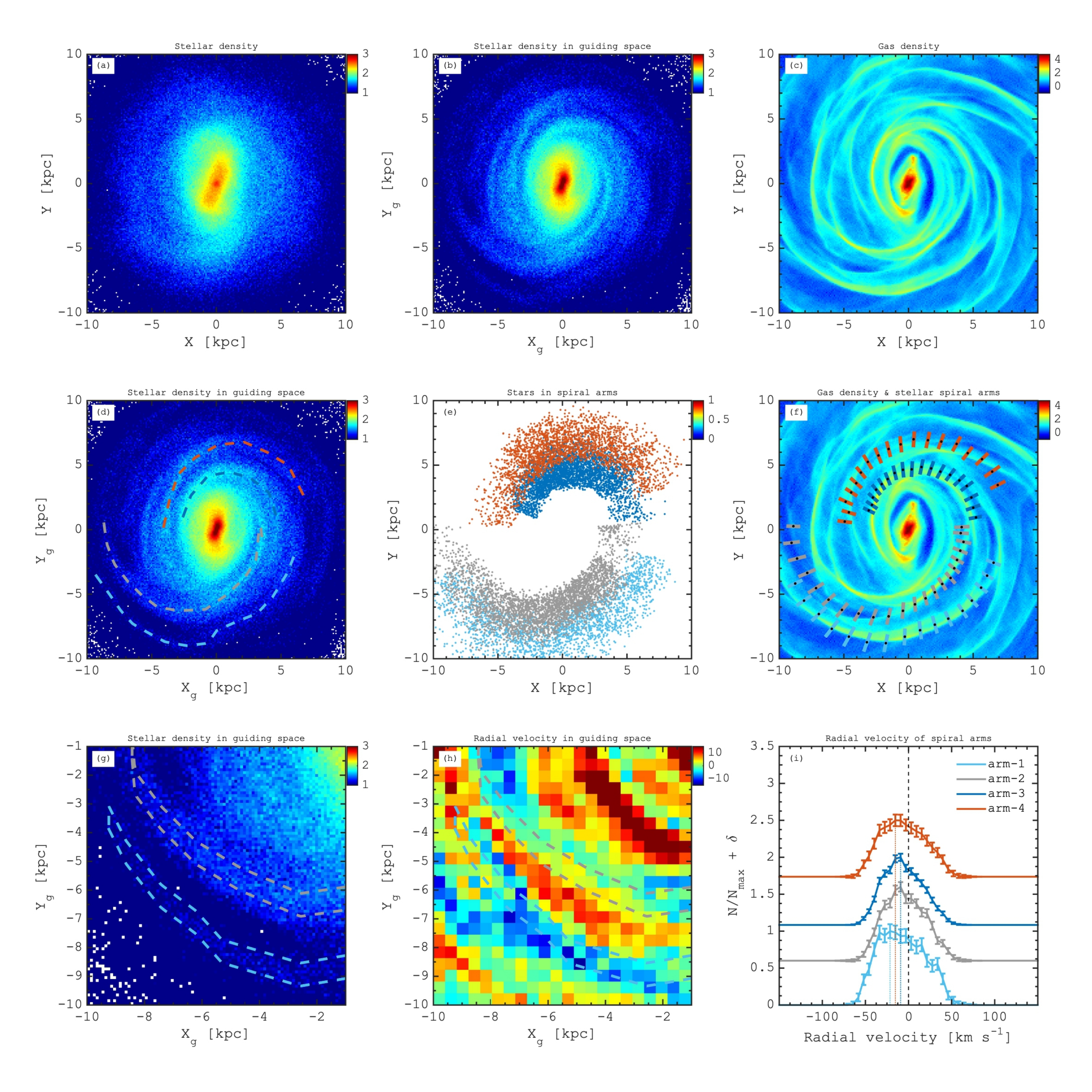}
\caption{Milky Way-type galaxy simulation~(Model I): behaviour of spiral arms. {\it (a), (b):} Face-on stellar surface density in spatial coordinates and in the guiding phase space respectively. {\it (c):} Face-on surface density of gas. {\it (d):} Same as {\it b} but with the colour lines depicting the position of four spiral arms features. {\it (e):} Spatial distribution of star particles associated with the spiral-arm features detected  guiding space. Colours of points correspond to the colour of lines in {\it (d)}. {\it (f):} Gas surface density distribution with a position of stellar spiral arms overplotted. {\it (g):} Zoom-in of the guiding space containing two prominent spiral arms highlighted by the coloured lines. {\it (h):} Distribution of the mean radial velocity in the zoomed-in region of the guiding space where the contours show the spiral arm location. {\it (i):} Distribution of the radial velocities of stars in four spiral arms identified in frames {\it (d)} and {\it (e)} where the spiral arms stars have skewed distribution with negative mean values. We note that the errors in the radial velocity measurements are determined from the scatter around the mean value, using the standard deviation of the measurements. The kinematical features presented here confirm that the structures with similar kinematics found in the guiding space of the Milky Way are the spiral arms.}\label{fig::sph2}
\end{figure*}

Model I is an $N$-body/hydrodynamical simulation of the stellar-gaseous galactic disc of the Milky Way-type galaxy. The gaseous subsystem is modeled as Smoothed Particle Hydrodynamics Particles~(SPH), and the dynamics of the stellar disc is modeled by using a direct particle-particle integration scheme. To avoid the change of the bar size due to possible angular momentum exchange between the disc and the halo/bulge system resulting in a slowdown of the bar, we use a rigid galactic potential for spherical components. The importance of a live DM halo has been noticed in several works, where the angular momentum exchange between the halo and the disc plays an important role in increasing the bar length over time~\citep{2000ApJ...543..704D,2003MNRAS.341.1179A}. However, the impact of live DM halo on the formation and evolution of the spiral arms is not evident. Rigid halo simulations reproduce well radial migration of stars~\citep[e.g.,][]{2012MNRAS.421.1529G,2013A&A...553A..77G, 2014ApJ...794..173V} and are in agreement with theories of spiral structure formation~\citep[e.g.,][]{2011MNRAS.410.1637S,2012MNRAS.427.1983K,2012ARep...56...16K,2013ApJ...766...34D, 2019MNRAS.489..116S}. Thereby we are confident that, at least on the qualitative level, our Model I is a reasonable guide for understanding the MW spiral arms kinematics. In our simulation, the number of particles is $2^{21}$ and $2^{22}$ for stars and gas respectively, and the softening length is about $40$~pc for both components. The generation of the model initial equilibrium follows an iterative method of the Jeans equation solution~\citep{2012MNRAS.427.1983K}. The simulation was performed with a direct GPU $N$-body/SPH code which provides a very high accuracy of the numerical integration~\citep{SPH}. We focus our analysis on a single snapshot of the evolution at $1.3$~Gyr when an extended multi-arm tightly-wound spiral structure has formed in the model.

In Fig.~\ref{fig::sph2} we show this simulated multi-arm spiral galaxy where the spirals are mostly seen in the gas density distribution while the stellar density is smooth and does not reveal any clear signatures of the spirals arms. Then, similar to the \Gaia~DR~2 data analysis, we calculated the stellar density distribution in the guiding $\rm (X_g,Y_g)$ phase-space coordinates where the density map~(see Fig.~\ref{fig::sph2}a) demonstrates a few large-scale features similar to those we discovered in \Gaia~DR~2. To proceed further with simulated data, we selected stars along four major features in guiding phase-space and show the face on distribution of the stars associated with these features~(see Fig.~\ref{fig::sph2}e). These stars, taken in narrow regions in the guiding space, cover a large area in the disc plane, with some overlap, and in order to constrain the location of structures we provide a 1D-gaussian fitting of the structure's location and their width in $\rm (X,Y)$ coordinates, similar to those we use for the \Gaia data. In Fig.~\ref{fig::sph2}f we compare the derived stellar density structures location with the gas density distribution. Note that the stellar spiral arms mostly co-exist with the gaseous ones. Because the gas component does not have the epicyclic motion, which leads to a large variation in the velocities for stars, there is a narrow structure in the gas seen along the stellar arm. However, at the edge of spiral arms, a spatial offset between gaseous and stellar spirals is clearly visible which is expected due to hydrodynamical instabilities of the shocks in a weak spiral potential~\citep{2004MNRAS.349..270W,2011AstL...37..563K}. Once we select stars in the guiding space features~(see Fig.~\ref{fig::sph2}g) we calculate the radial velocity distribution~(see Fig.~\ref{fig::sph2}h). For this model and for the the observed SDS structures error bars are determined using the standard deviation of the measurements. The distributions show that the stars associated with spiral arms move inwards~(negative radial velocity). This systematic motion explains the mean radial velocity of stars in the spiral arms and their skewed distribution profile~(see Fig.~\ref{fig::sph2}h). This result is further evidence that stellar density structures with similar kinematics~(SDS~1, 3, 4, 6) correspond to the spiral arms of the the Milky Way. In Model I, spiral structure is a slowly rotating pattern with a corotation outside $10$ kpc, where a systematic outward motion (positive radial velocity) of stars is observed in agreement with the density wave theory. 

\begin{figure*}[t!]
\includegraphics[width=1\linewidth]{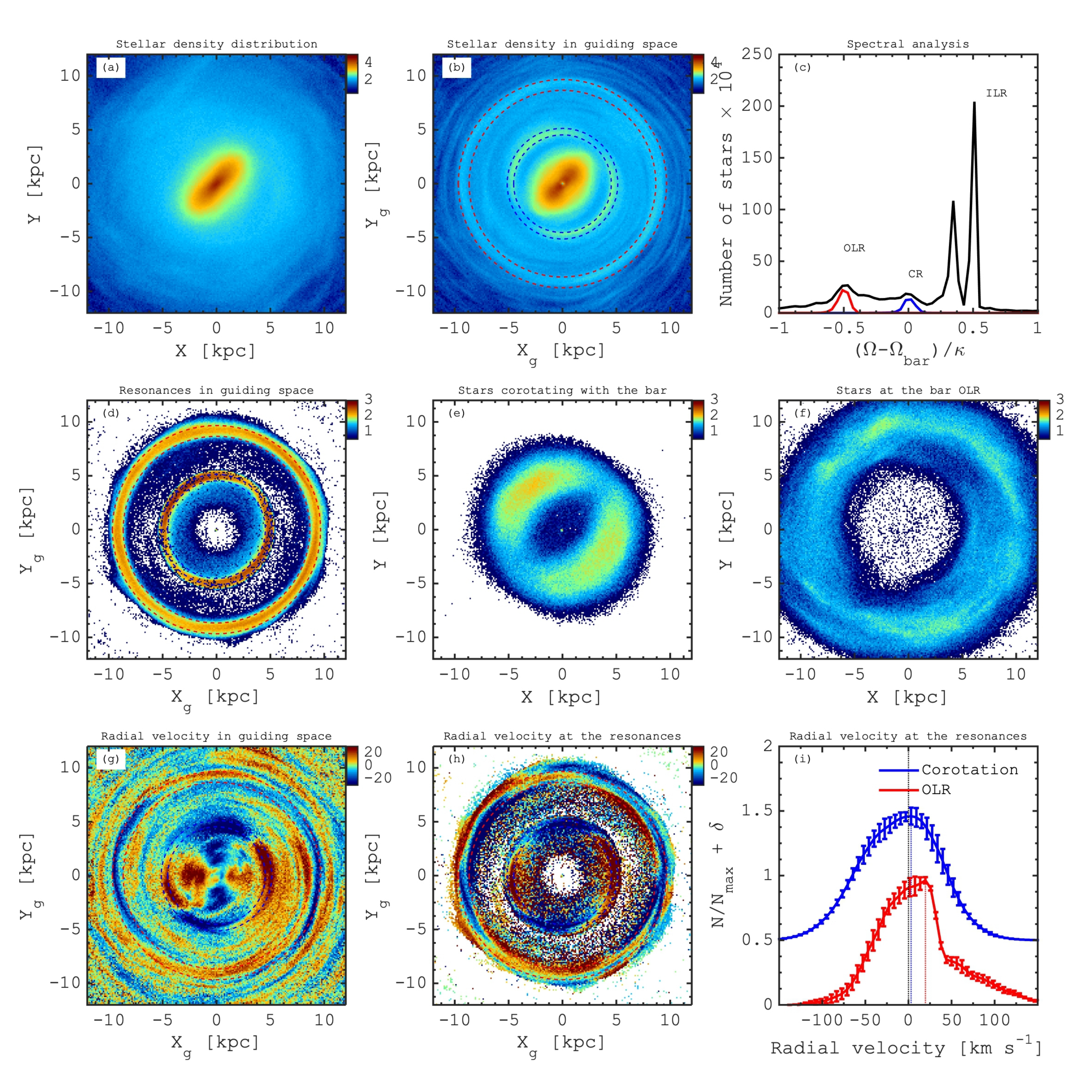}
\caption{Milky Way-type galaxy simulation~(Model II): structures near the bar resonances. {\it (a):} Face-on map of the stellar surface density. {\it (b):} Stellar surface density in the guiding space. Two distinct quasi-circular overdensities appear close to the end of the bar and in the outer disc highlighted by blue and red circles respectively. {\it (c):} Spectral analysis of orbits of stars where the fraction of stars with different $\rm F = (\Omega - \Omega_{\rm bar}) / \kappa$ ratio is shown for all disc stars~(black) and stars in between blue and red circles in frame {\it (b)}. Frequency ratios of $-0.5$, $0$ and $0.5$ correspond to the main resonances of the bar -- outer Lindblad, corotation and inner Lindblad resonances respectively. {\it (d):} Guiding space density distribution of stars at the bar corotation ($|F|<0.05$) and at the OLR~($|F +0.5|<0.05$). {\it (e):} Density distribution of stars co-rotating with the bar in spatial coordinates. {\it (f):} Density distribution of stars at the bar OLR in spatial coordinates. {\it (g):} Mean radial velocity distribution of all stars in the guiding space. {\it (h):} Mean radial velocity distribution of stars at the bar corotation and the OLR in the guiding space. {\it (i):} Distributions of the radial velocities of stars at the bar corotation~(blue) and at the OLR~(red).}\label{fig::run6}
\end{figure*}

\subsection{Model II: barred galaxy model}\label{app2b}
Model II is a purely collisionless $N$-body simulation of a composite disc galaxy with live dark matter halo with a total number of particles of $10^7$ for stars and $0.5\times10^7$ for dark matter. For the $N$-body system integration, we used our parallel version of the TREE-GRAPE code with multithread usage under the SSE and AVX instructions. The tolerance parameter of the tree-code is $\Theta = 0.7$, and for smoothing, we use a Plummer potential with $50$~pc. For the time integration, we used a leapfrog integrator with a fixed step size of $0.2$~Myr. This simulation has been previously used to reproduce the morphology of the metal-rich and metal-poor stellar populations in the Milky Way bulge~\citep{2017A&A...607L...4F,2018A&A...616A.180F}, to understand the origin of the phase-space spirals in isolated Milky Way-type galaxies~\citep{2019A&A...622L...6K} and to explain the ridges in $R-V_\phi$-the plane of the Milky Way~\citep{2019MNRAS.485L.104M,2019MNRAS.488.3324F}. In this work, we refer in our analysis to the snapshot at $4$~Gyr from the beginning of evolution, when no prominent spiral structure is seen, and the bar is the dominant non-axisymmetric structure perturbing the phase-space of the simulated galaxy~(see Fig.~\ref{fig::run6}a). 

In Fig.~\ref{fig::run6} we show the density distribution of stars in both spatial and guiding space where two prominent quasi-circular distinct features appear in the guiding space. The motion of a star in a galactic plane is described by radial~($\kappa$ -- epicyclic frequency) and angular oscillation~($\Omega$ is the rotation around the galactic centre). To confirm that these overdensities are related to the bar resonances we measure the orbital frequencies~($\Omega, \kappa$)~\citep{1982ApJ...252..308B,2007MNRAS.379.1155C,2018A&A...616A..86H} of all particles over a $1$~Gyr period around $4$~Gyr. Each of the bar resonances is characterized by a certain value of $F = (\Omega - \Omega_{\rm bar})/\kappa$, e.g. for the bar corotation $F=0$, and for inner~(ILR) and outer Lindblad~(OLR) resonances $F=-0.5$ and $0.5$ respectively. In Fig.~\ref{fig::run6}c the overall distribution peaks at the ILR, which is expected for a galaxy with a strong bar while the OLR is rather weak but slightly stronger than the corotation. Coloured lines show the frequency ratios for stars associated with two prominent features in the guiding phase-space where the inner ring-like structure frequency ratio peaks at corotation~(blue) and the outer ring consists in the OLR stars~(red). Therefore the two features we found in the guiding space of the barred galaxy correspond to the corotation and the OLR, or, in other words, the main bar resonances are imprinted in the guiding space as large-scale overdensities. Stellar density distributions and radial velocity maps of stars at the main bar resonances in guiding and spatial scales are presented in Fig.~\ref{fig::run6}d-h where stars that are trapped at the bar resonances cover a large area across the disc. The radial velocity distribution along the bar corotation region is roughly symmetric and the variations of the mean value are caused by sequential gain and loss of angular momentum while there are no net changes in angular momentum over a long period of time~\citep{2007MNRAS.379.1155C}. At the bar OLR stars librate around the $2:1$ closed periodic orbits that correspond to the $\rm x_1(1)$ and $\rm x_1(2)$ families which contribute to creating the asymmetry in radial velocity along the OLR~\citep{2019MNRAS.488.3324F}. These changes of the radial velocities are imprinted in the radial velocity distribution for stars at the resonances~(see Fig.~\ref{fig::run6}i), in particular, for a disc region similar to that available in the Milky Way~($\rm |X|<5$~kpc and $\rm Y<0$), the mean radial velocity is negative and the distribution spreads over a wide range of values. At the OLR, the radial velocity distribution is dominated by stars on $\rm x_1(1)$ orbits which have positive radial velocities and, on average this OLR patch moves outwards~(positive radial velocity). In Fig.~\ref{fig::run6}i error bars in the radial velocity distributions are from comparing $20$~simulation snapshots within a $\pm0.5$~Gyr range, using the standard deviation of the measurements. Both phase-space and kinematical features we found for stars at the resonances are in qualitative agreement with those we discovered in the Milky Way disc by analyzing the \Gaia~DR~2~(see Fig.\ref{fig::fig2}). Therefore, we confirm that the SDS~2 is likely to be associated with the corotation while the SDS~5 corresponds to the OLR of the bar.

\section{Guiding space analysis}\label{app1}

\begin{figure*}[t!]
\includegraphics[width=1\linewidth]{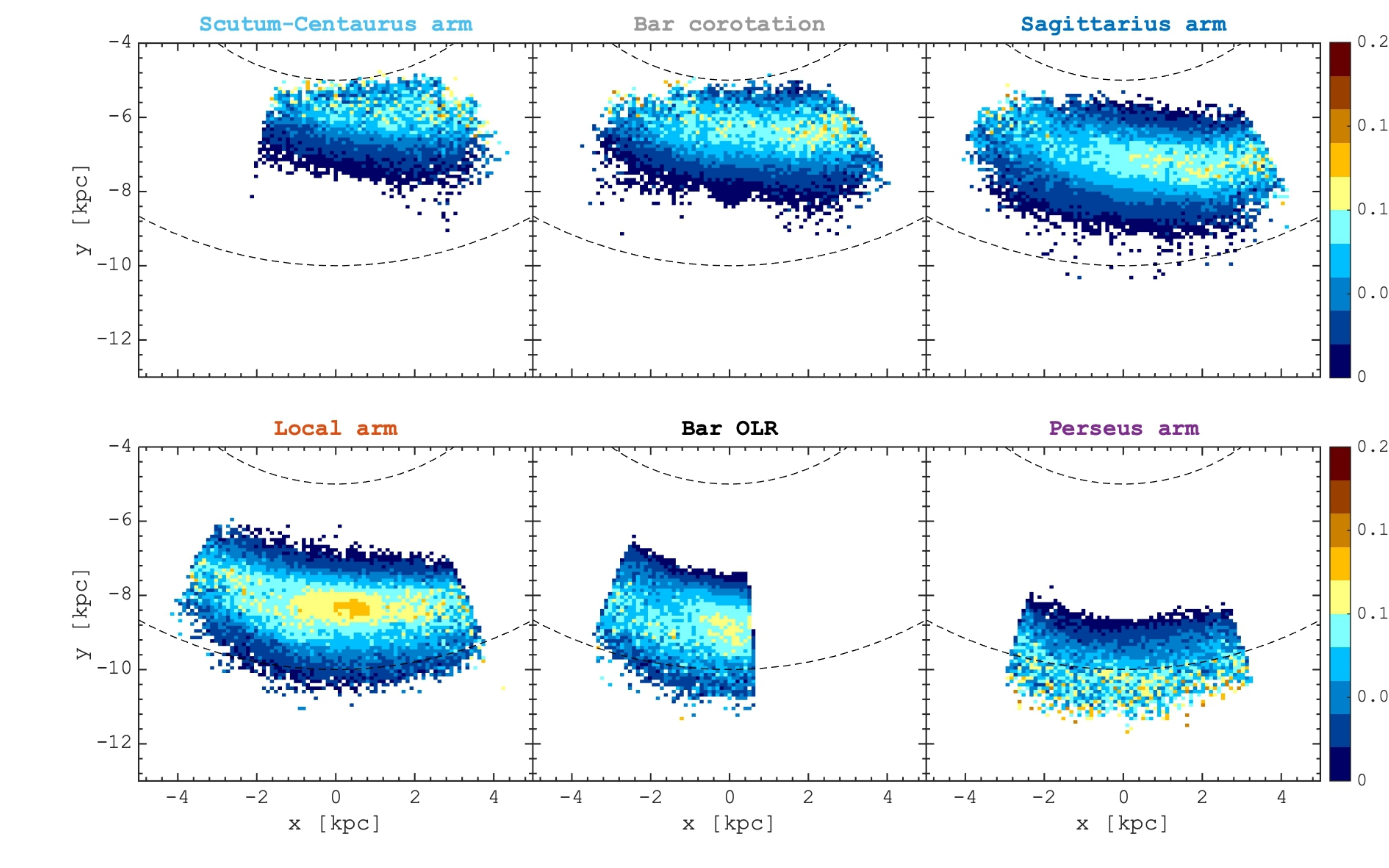}
\caption{Amplitude of the stellar density structures causing the large-scale overdensities in the Milky Way phase-space in GRV2. Each frame depicts a fraction of stars corresponding to different guiding space features in the entire \Gaia~DR~2 footprint. }\label{fig::fig4}
\end{figure*}

To explore the physical properties of the features we find in the guiding coordinates space of the Milky Way, we first select stars associated to these structures in the guiding space and plot their distribution in the physical space. For each guiding space overdensity we select stars which have guiding radius not further than 200 pc from the mean value for a given azimuthal position. In Fig.~\ref{fig::fig4}, we show the $(X,Y)$ maps of the fraction of these selected stars, for each structure, relative to the total number of stars at the same location. These density maps clearly show that the physical location of the recovered structures spans a very large area from the inner $5$~kpc region to the outer disc at $11$~kpc. A notable feature is that the amplitude of each structure is about $5-12$\% relative to the total stellar density at a given location.

To place all structures in the Galactic plane, we estimate their mean location from Fig.~\ref{fig::fig4} by using Gaussian fitting. In particular, we fit the density distributions~(see Fig.~\ref{fig::fig4}) along the galactocentric radial direction with 1D Gaussian models. The result of this procedure is overplotted in Fig.~\ref{fig::fig4}, where we also show the width of the structures, calculated as the $1$-sigma dispersion in positions along the radial direction. This width slightly varies along the SDS, and for the different structures, changing from $0.5$~kpc, for the innermost spirals, up to $\approx2$~kpc for the Perseus arm beyond the Solar vicinity. The thickness may seem significant, but it must be noted that no selection on age has been made, and our sample, therefore, includes stars with different kinematics, some of them having large epicyclic motions. Large empirical scatter between different tracers of spiral arms has been found also in nearby spiral galaxies~\citep{2008AJ....136.2872T,  2011ApJ...735..101F, 2013ApJ...779...42S} making spiral arms rather wide for stars of different populations.

\end{appendix}

\bibliographystyle{aa}
\bibliography{references}

\end{document}